\documentclass[twocolumn,nofootinbib,aps,showpacs,pra,preprintnumbers]{revtex4-1}

\usepackage{graphicx}
\usepackage{dcolumn}
\usepackage{epsfig}
\usepackage{epstopdf}
\usepackage{amssymb}
\usepackage{amsfonts}
\usepackage{xcolor}
\usepackage{float}
\usepackage{amsmath}
\usepackage{hyperref}
\usepackage{physics}
\usepackage{comment}
\usepackage[caption=false]{subfig}
\usepackage{soul}
\usepackage[bottom]{footmisc}

\hypersetup{
	colorlinks = true,
	linkcolor = blue,
	citecolor = blue,
	urlcolor = blue
}
\allowdisplaybreaks
\setstcolor{red}

\begin{document}

\title{Transition frequency measurement of highly excited Rydberg states of $^{87}Rb$ for a wide range of principal quantum numbers}

\author{Silpa B S}
\author{Shovan Kanti Barik}
\author{Saptarishi Chaudhuri}
\author{Sanjukta Roy}
\email{sanjukta@rri.res.in}

\affiliation{Raman Research Institute, C. V. Raman Avenue, Sadashivanagar, Bangalore 560080, India}

\begin{abstract}
We report our measurements of the absolute transition frequencies of $5P_{3/2}, F=3$ to $nS$ and $nD$ Rydberg states of $^{87}Rb$ with high principal quantum numbers in a wide range of values (n =45 -124). The measurements were performed using Rydberg Electromagnetically Induced Transparency (EIT) in ladder-type three-level systems. We measure the transition frequencies with an accuracy of $\le$ 2 MHz. We determine the values of the Rydberg-Ritz parameter for $^{87}Rb$ from our experimental measurements of the transition frequencies. Our measurements of the absolute transition frequencies of the highly excited Rydberg states would be useful for diverse applications in quantum information processing, quantum simulation and quantum sensing with Rydberg atoms.
\end{abstract}
\maketitle

\section{Introduction}
Rydberg atoms offer a unique platform for the reliable implementation of scalable quantum information processing due to their strong inter-atomic dipole-dipole interactions \cite{Saffman2010}. The highly controllable long-range interactions between Rydberg atoms make them excellent atomic building blocks for quantum technologies \cite{Adams_2019} and scalable quantum information networks \cite{Kimble_2008} as well as single photon source for secure quantum communications \cite{Pfau_single_photon}. Atoms excited to Rydberg states with high principal quantum numbers $n$ have exaggerated properties such as strong dipole-dipole interaction ($\propto n^4$), large values of polarizability ($\propto n^7$) and longer life-times scaling as $n^3$. These make highly excited Rydberg atoms ideal candidates for quantum information processing \cite{Saffman2010}, precision electrometry \cite{Haroche_electrometer}, digital communication\cite{2019_Song} as well as offering a versatile platform for novel atomic sensor technologies \cite{Fan_2015}. 
\par
The knowledge of the absolute transition frequencies to the highly excited Rydberg states is required to access their extraordinary characteristics for experiments. Rydberg Electromagnetically induced transparency (EIT) \cite{2007_Mohapatra, 2022_Opt_Ryd_EIT, 2017_Cheng, 2021_Xue, 2021_Ji, 2018_Zhang} is one of the most efficient and non-destructive ways to detect atoms in the Rydberg states. In a closed three-level atomic system, EIT occurs due to the destructive quantum interference between the transition probabilities. This results in cancellation of absorption of the probe beam and producing transmission in the atomic medium on an atomic resonance \cite{EIT_Review_Marangos}. In addition, EIT has important applications in slowing of light \cite{1999_Hau}, light storage in atomic medium \cite{2001_Lukin} and quantum memory \cite{Ma_2017}. In this work, we utilize Rydberg EIT in a three-level ladder-type system of $^{87} Rb$ for the measurement of the transition frequencies to highly excited nS and nD Rydberg states.
\par
 In this work, we report our comprehensive measurements of the absolute transition frequencies of the $5P_{3/2}, F=3 \rightarrow nS_{1/2}$, $5P_{3/2}, F=3 \rightarrow nD_{3/2}$ and $5P_{3/2}, F=3 \rightarrow nD_{5/2}$ transitions to the Rydberg states of $^{87}Rb$ with high principal quantum number $(n=45 -124)$ with an accuracy of  $ \le $ 2 MHz. The Van der Waals interaction between Rydberg atoms varies as $n^{11}$ resulting in higher interatomic interaction for atoms in highly excited Rydberg states with large values of principal quantum number. Our measurements would enable accurate identification of highly excited Rydberg states with large values of principal quantum number n having remarkably long lifetimes ($\propto n^3 \sim 100 \mu s$), large sizes ($\sim 1 \mu m$) and large dipole moments ($\propto n^2$) and thereby enhanced strengths of inter-atomic interaction. This long-range dipolar interaction forms the basis for the implementation of quantum entanglement between Rydberg atoms via Rydberg blockade \cite{2010_Browaeys} and realisation of fast quantum gates \cite{2010_PRL_Saffman}. Previously, measurement of Rydberg transition frequencies using Doppler-free two-photon spectroscopy \cite{1979_CJP} were performed, identification of $nF_{7/2}$ and $nS_{1/2}$ Rydberg states of $^{85}Rb$  \cite{Johnson_2010, 1985_Weber} were done, measurement of selected Rydberg levels between n=19-92 were carried out \cite{2011_Mack, Watanabe_2017}. In our work reported in this paper, we present a systematic experimental study and tabulate the Rydberg transition frequencies in the range n=45-124 with high precision ($\le 2 $ MHz). This will enable researchers to conveniently access these highly excited Rydberg levels for a myriad of applications.
 \par
 In the following sections, we present a brief overview of Rydberg atoms and Quantum defects (Sec.\ref{sec:Theory}) and detailed description of the experimental setup and methods for measurement of the transition frequencies (Sec.\ref{sec:Expt_setup}). In Sec.\ref{sec:Measurements}, the experimental results for the variation of the absolute transition frequencies of the Rydberg states and the corresponding variation of the fine-structure splitting as well as the transition strengths with the principal quantum number n are presented. The values of the quantum defects (Rydberg-Ritz parameters) obtained from the experimental measurements are also presented. Finally, we present some concluding remarks and future perspectives.
 
\section{Rydberg atoms: Quantum defects}\label{sec:Theory}
 Rydberg atoms are atoms with the outer electron excited to an electronic orbit far away from the nucleus. Due to the screening of the charge of the nucleus by the electrons in the inner shells, atoms in the Rydberg state can be considered as hydrogen-like atoms. The energy of the atoms in the Rydberg state with principal quantum number $n$, orbital angular momentum quantum number $l$, and total angular momentum quantum number $j$ with respect to the ionisation limit $E_i$ is given by \cite{gallagher_1994, Greene_RMP}:
 
\begin{equation} 
\label{eq:QD}
E_{nlj}=E_i-\frac{R_{Rb}}{[n-\delta_{nlj}]^2} 
\end{equation} 

where $\delta_{nlj}$ is the quantum defect that occurs because, for alkali atoms with lower orbital angular momentum quantum number, the wavefunction of the valence electron penetrates the ionic core thereby reducing the screening of the Coulomb potential. This results in the enhancement of the effective charge seen by the electron in the Rydberg state and hence the lowering of its energy quantified by the quantum defect. The Rydberg constant for $^{87}Rb$, calculated by correcting for the reduced mass of the electron, is given by:

\begin{equation}
\label{eq:Rydberg_const}
\begin{split}
\small
R_{Rb} = {\frac{m_{Rb}}{m_{e} +m_{Rb}}} R_y= h \cdot 3.28982119466 (2) \times 10^{15} Hz 
\end{split}
\end{equation}

where $R_y$ is the Rydberg constant (CODATA) is given by:
\begin{equation}
\label{eq:Rydberg_const}
\begin{split}
\small
R_{y} = \frac{m_e e^4}{8 {\epsilon_0}^2h^{3}c}= 10973731.568160(21)  m^{-1},
\end{split}
\end{equation}
$m_e$ and $e$ are the mass and charge of the electron respectively, $\epsilon_0$ is the permittivity of free space and h is the Planck's constant. 

Quantum defect of an  nlj state can be expressed as  \cite{gallagher_1994}:
 \begin{equation}
     \delta_{nlj}=\delta_0 +\frac{\delta_2}{(n-\delta_0)^2}+\frac{\delta_4}{(n-\delta_0)^4}+\frac{\delta_6}{(n-\delta_0)^6}+........
\end{equation}
where, $\delta_k  $ are the Rydberg-Ritz parameters. 
For large values of the principal quantum number n, the quantum defects $\delta_{nlj}$ have low dependence on n and can be written approximately as:

\begin{equation}
\delta_{nlj} \approx \delta_0 + \frac{\delta_2}{(n-\delta_0)^2}
\label{approx_QD}
\end{equation}

\par
Rydberg-Ritz parameters are generally obtained from the fine-structure levels of the Rydberg states. The hyperfine splitting is negligible for the nD states and it is significant only for the low-lying nS states due to the penetration of the valence electron wavefunction into the nucleus. Earlier measurements of Rydberg-Ritz parameters are reported in \cite{2003_Li, 2021_Results_physics, Sanguinetti_2009, 2011_Mack, 2019_Raithel, 2013_Spreeuw}.

The scaling law for the hyperfine splitting of the $nS$ Rydberg levels is given by \cite{2013_Spreeuw}:
\begin{equation}
\Delta_{hfs, F=2} - \Delta_{hfs, F=1} = 37.1(2) \, \text{GHz} \times \, {n^{*}}^{-3}
\label{hfs}
\end{equation}
where, $n^{*}$ is the effective principal quantum number given by $n^{*}= n-\delta_{nlj}$. We used this most recent measurement of the hyperfine splitting of the nS states in $^{87}Rb$ to estimate the correction to our measured transition frequencies.
\par
We measure the transition frequencies only for $nS_{1/2}, F=2$ hyperfine states since the excitation from $5P_{3/2}, F'=3$ state to the $nS_{1/2}, F=1$ is dipole forbidden. Using Eqn.{\ref{hfs2}} and Eqn.{\ref{hfs3}} from Appendix \ref{sec_hfs}, the hyperfine shift of the $nS_{1/2}, F=2$ Rydberg state is given by:
\begin{equation}
\Delta_{hfs, F=2} = \frac{3}{4} A= \frac{3}{8} 37.1(2)\, \text{GHz} \times \, {n^{*}}^{-3}
\label{hfs4}
\end{equation}
We subtract the hyperfine shift obtained from Eqn.{\ref{hfs4}} from the measured transition frequencies of the $nS_{1/2} $, F=2 Rydberg states to find the corresponding quantum defects shown in Table \ref{tab:QD}.

\section{Experimental setup and methods}
\label{sec:Expt_setup}
\begin{figure}[h!]
  \centering
    \includegraphics[scale = 0.23]{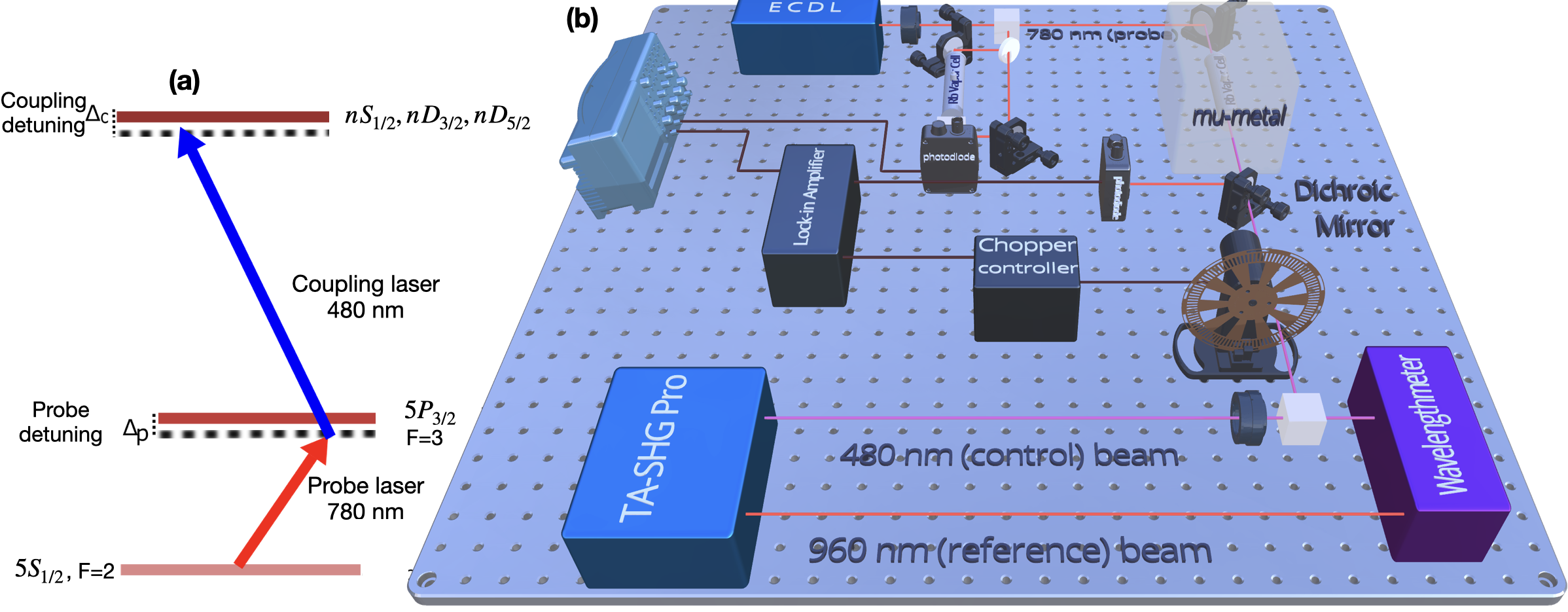}
  \caption{(a) Energy-level diagram of the ladder-type three-level atomic state configuration for the observation of Rydberg electromagnetically induced transparency. (b) Schematic of the experimental set-up for fine structure spectroscopy of Rydberg states using Rydberg EIT. } 
  \label{Expt_setup}
\end{figure}

A schematic of the experimental setup is shown in Fig.{\ref{Expt_setup}}. The probe laser was derived from an external cavity diode laser (ECDL) from Toptica (DL100) and was tuned to 780 nm which is the transition wavelength of the D2 atomic transition ($5S_{1/2}, F=2  \rightarrow 5P_{3/2}, F'=3$)  of Rb. The coupling laser was derived from a frequency-doubled tunable laser (Toptica TA SHG-pro) and was tuned from $5P_{3/2}, F'=3 \rightarrow nS_{1/2},nD_{3/2},nD_{5/2}$ around 480 nm by tuning the Master laser operating at 960 nm. We measured the frequencies of the Master laser and the frequency doubled Coupling laser using a commercial Fizeau interferometer-based wavelength meter (HighFinesse WS8-2) with an absolute accuracy of 2 MHz and a frequency resolution of 200 kHz. The wavelength meter was kept in a thermally insulated box to avoid drifts of the interferometer signal due to ambient thermal fluctuations. The wavelength meter was frequently calibrated using the known frequency of a laser locked to the $5S_{1/2}, F=2  \rightarrow 5P_{3/2} F'=3$ transition with long term stability of $\approx$ 400 kHz (refer to Appendix \ref{freq_stability}). The coupling laser frequency was varied by tuning the frequency of the Master laser of the TA SHG-pro. 
\par
A collimated probe beam having an FWHM of 0.8 mm was aligned through a Rb vapour cell of length 10 cm as shown in Fig.{\ref{Expt_setup}}. The Rb vapour cell was enclosed in a $\mu$-metal magnetic shielding to prevent the influence of any ambient stray magnetic field on the atoms. The coupling beam having FWHM of 0.7 mm was aligned through the vapour cell overlapping with the probe beam in the counter-propagating direction to minimise the Doppler mismatch effect as discussed in detail later. 
\par
The probe was locked to the $5S_{1/2},F=2  \rightarrow 5P_{3/2}, F'= 3$ transition with a frequency uncertainty of $\approx$ 400 kHz. The coupling laser was scanned across the Rydberg EIT resonance and the wavelength of the coupling beam was recorded simultaneously using the wavelength meter. A typical probe transmission signal at a Rydberg EIT resonance for $5P_{3/2}, F'=3 \rightarrow 50D_{5/2}$ Rydberg transition with the coupling laser scanning across the resonances is shown in Fig.{\ref{fig:diff} (a).  This method provided the measurements of the absolute transition frequencies of the $5P_{3/2}, F'=3 \rightarrow  nS_{1/2}, nD_{3/2}, nD_{5/2}$ transitions. For each of the Rydberg EIT resonances, the signal was recorded 100 times and thereafter averaged to estimate the peak frequency of the resonance.
\par
We could measure the absolute transition frequencies of the $5P_{3/2}, F' = 3 \rightarrow nS_{1/2}$ and $5P_{3/2}, F'=3 \rightarrow nD_{3/2}, nD_{5/2}$ transitions up to n=80 without using lock-in detection with a good signal to noise ratio (SNR) $>$ 10 of the Rydberg EIT signal. From n=81 onwards, we used lock-in detection to record the Rydberg EIT resonances to maintain a good SNR for accurate determination of the coupling frequency corresponding to the peak of the Rydberg EIT resonance signal. The lock-in detection was done using a mechanical chopper in the path of the coupling beam and a lock-in amplifier (Stanford Research systems, SR-850) as shown in Fig.\ref{Expt_setup}. The chopper was used to modulate the intensity of the coupling beam by periodically blocking the coupling beam at a frequency of $\sim$ 4 kHz and the photodiode signal of the probe beam was recorded with a good SNR using a lock-in amplifier synchronized using the reference signal from the chopper as shown in Fig.{\ref{fig:diff}(b).
\par
Another method used for recording the Rydberg EIT signals was lock-in detection via modulating the frequency of the probe laser. The coupling laser was kept at a frequency corresponding to the measured transition frequency of the Rydberg state using the method described above and the probe was scanned across the EIT resonance. The signal was recorded using lock-in detection by modulating the frequency of the probe beam at a modulation frequency of $\approx$ 4 kHz, with small modulation amplitude. The second harmonic of the Rydberg EIT signal was recorded simultaneously with the saturation absorption spectroscopy signals of $^{87}Rb$ to calibrate the frequency detuning of the probe beam. The probe detuning was converted to the coupling detuning using Eqn.\ref{Doppler_mismatch}. The Rydberg EIT signal obtained using this method is shown in Fig.{\ref{fig:diff}(c). This method is useful for studying the characteristics of the Rydberg EIT signals with long-term stability without causing any disturbance to the experimental system due to the mechanical vibration caused by the chopper. The data shown in Fig.\ref{fig:nS_freq2} and Fig. \ref{fig:nD_freq3} were recorded using this method.
\par
By comparing the Rydberg EIT resonances with and without lock-in detection for n $<$ 81 Rydberg states, we observed no discernible shift in the peak frequency of the Rydberg EIT resonance as shown in Fig.\ref{fig:diff}. The difference in the peak frequencies obtained via fitting the Rydberg EIT signals with and without lock-in-detection was $\le$ 600 kHz. The probe beam power was kept at a low value of 4 $\mu W$ to minimise the power-broadening effect on the signal. The variation of the linewidth of the Rydberg EIT signal with the probe power is shown in  Fig.{\ref{EIT_width_probe_power}}. The typical width of the Rydberg EIT signals measured in our experiment was around 3 - 5 MHz as shown in Fig.{\ref{EIT_width_probe_power}}.
\par
The Rydberg EIT resonance signals recorded without lock-in-detection were fitted with a Lorentzian function to determine the coupling laser frequency corresponding to the peak of the signal. For fitting the signals recorded with lock-in detection accurately, we have used the asymmetric pseudo-Voigt function \cite{2008_Voigt} because this function takes care of the asymmetricity of the signal recorded using lock-in detection. For fitting the signals obtained using the lock-in detection via modulation of the probe frequency shown in Fig.\ref{fig:diff}(c), the second derivative of the asymmetric pseudo-Voigt function was used. The details of the fitting method are described in Appendix \ref{sec_fit}.

 \begin{figure}[h!]
\centering\includegraphics[scale=.5]{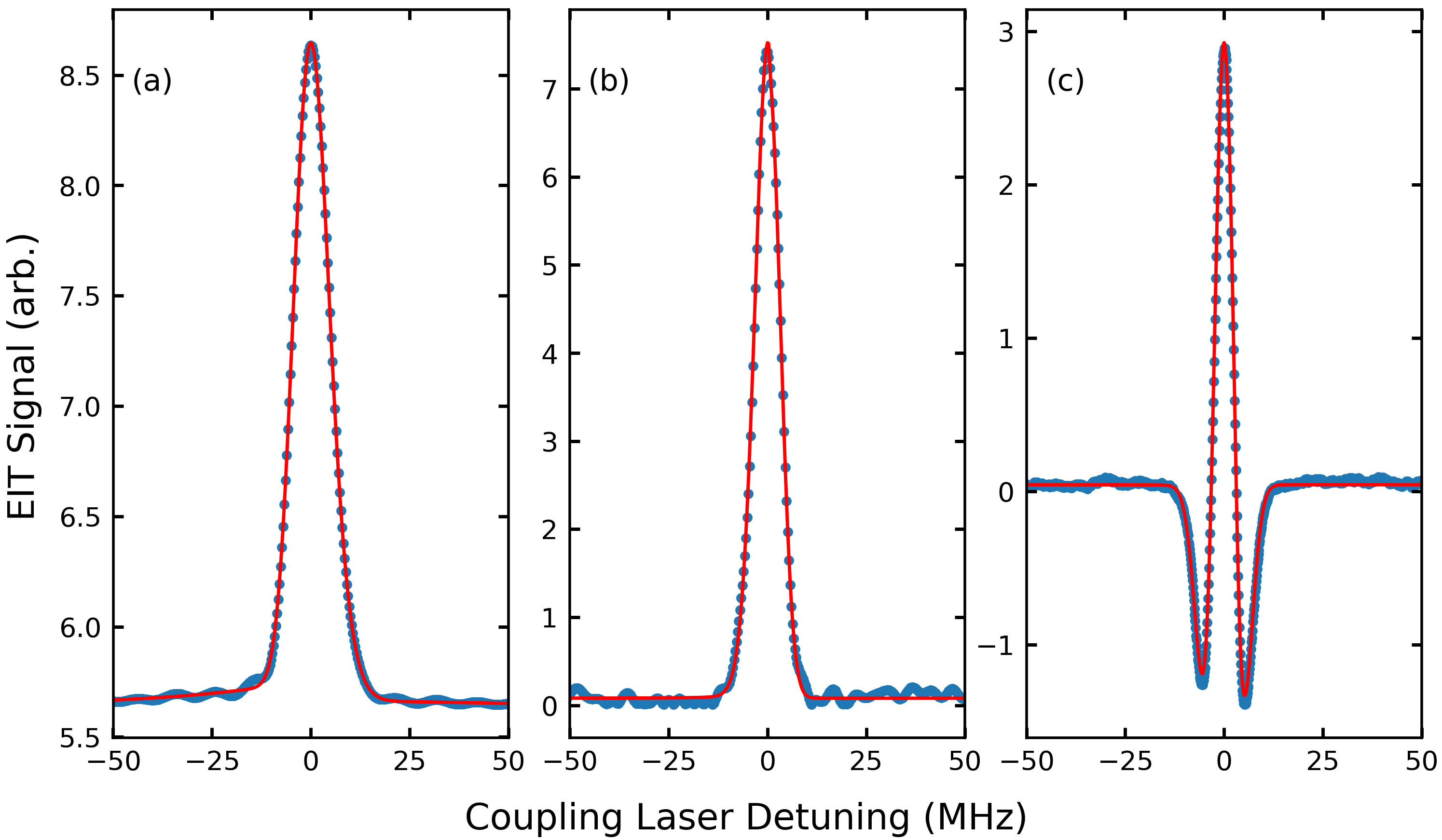}
\caption{Rydberg EIT signals recorded for the $50D_{3/2}$ Rydberg state (a) without lock-in detection (b) lock-in detection using modulated coupling beam with mechanical chopper (c) lock-in detection with modulated probe beam. The red solid lines are the fits to the experimental data. The difference in the values of the coupling laser frequencies obtained from fitting of each of these signals was $\le$ 600 kHz.}
\label{fig:diff}
\end{figure}

\par
A frequency detuning of $\Delta_p$ of the probe beam from the $5S_{1/2}, F=2  \rightarrow 5P_{3/2}, F'=3$ transition results in a corresponding shift of $\Delta_c$ of the coupling laser frequency at which the peak in the probe transmission occurs due to Doppler mismatch \cite{2007_Mohapatra, 2011_Mack}. In the ladder-type of a three-level system, the analytical expression for the two-photon absorption contains the term ($\Delta_p + \Delta_c$). Hence, the counter-propagating probe and the coupling beam having matching wavelengths can render the medium Doppler-free resulting in a narrow EIT signal. However, for wavelength mismatch between the probe and the coupling beams such as the three-level ladder system considered in this work, the mismatch of the probe and the coupling wavelengths (780 nm and 480 nm) leads to the Doppler mismatch effect \cite{Urvoy_2013}. For small values of $\Delta_p$ compared to the Doppler broadening, 
$\Delta_c$ is given by:
\begin{equation}
\Delta_c = (\omega_c/\omega_p)\Delta_p
\label{Doppler_mismatch}
\end{equation}
where $\omega_p$ and $\omega_c$ are the angular frequencies of the probe and the coupling beams on resonance to the $5S_{1/2}, F=2  \rightarrow 5P_{3/2}, F'=3$ and $5P_{3/2}, F'=3 \rightarrow nS, nD$ transitions respectively. To compensate for this effect, we measured the absolute frequency of the probe laser by comparison with the simultaneously recorded saturation absorption spectroscopy signal and corrected the corresponding coupling laser frequency according to Eqn.{\ref{Doppler_mismatch}}. The frequency uncertainty of $\approx$ 400 kHz of the probe laser corresponds to a frequency uncertainty $\approx$ 650 kHz of the coupling laser measured using the wavelength meter according to Eqn.{\ref{Doppler_mismatch}}. 
\par
The overall error arising due to the deviation of the Rydberg EIT signals from the Lorentzian line shape used for finding the frequency of the peak position of the signal, laser linewidths and other technical noise present in the experiment was estimated to be around 400 kHz. The error in determining the peak frequency from the fit of the Rydberg EIT signal recorded using lock-in-detection was estimated to be $\approx $ 600 kHz. The glass cell was kept at room temperature (300 K) and the resulting line-shift and broadening effects due to vapour pressure of $^{87}Rb$ accounting for $< 150 $ kHz. 

\section{Results and discussion}\label{sec:Measurements}

\begin{figure}[h!]
\centering\includegraphics[scale=.2]{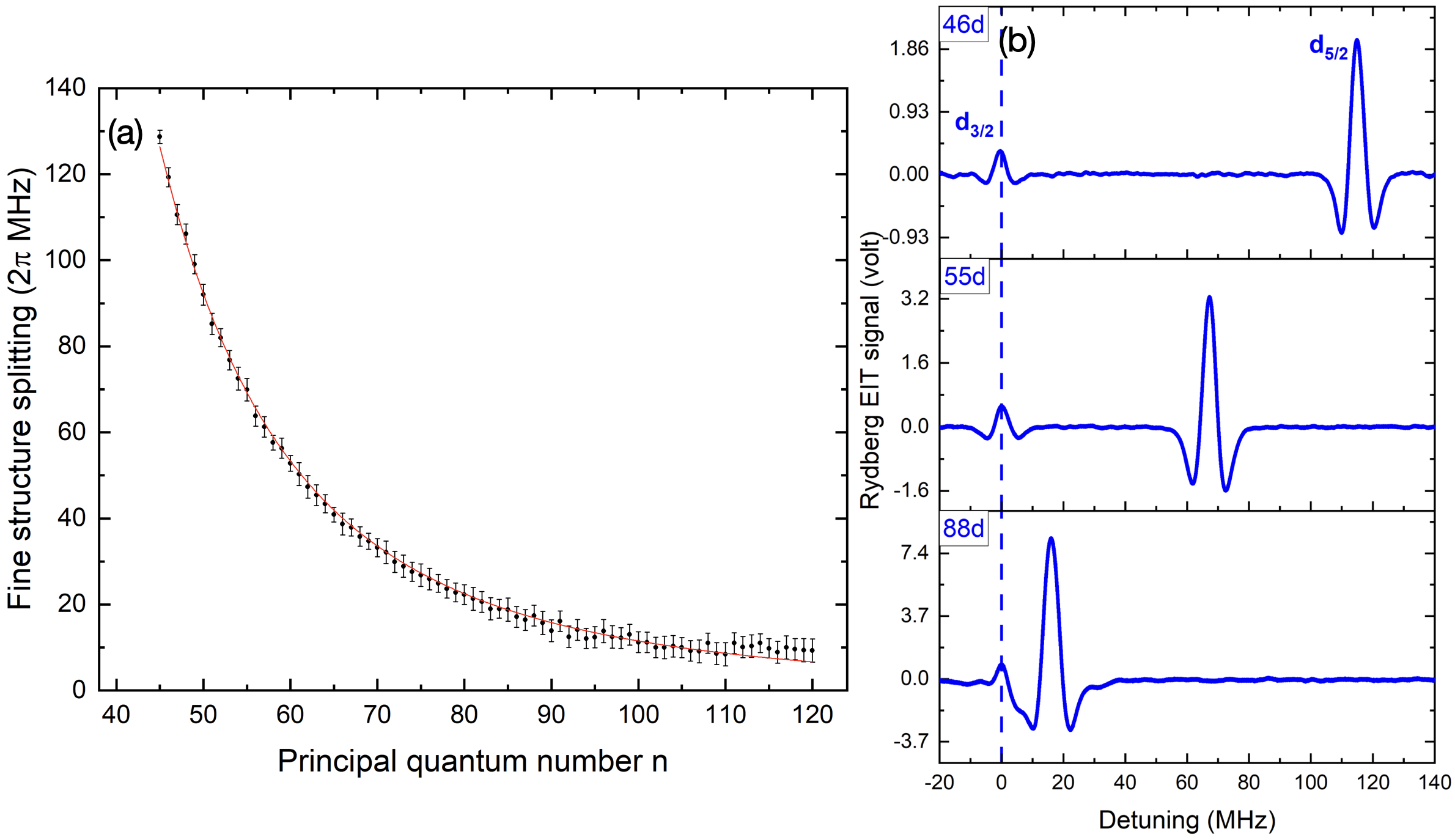}
\caption{(a) Variation of the fine structure splitting between the $nD_{3/2}$  and $nD_{5/2}$ Rydberg states with the principal quantum number n. The solid line is the fit to the data.(b) Fine structure splitting for three different principal quantum numbers n=46, n=55 and n=88 depicting the decrease in the splitting for higher Rydberg states.}
\label{fig:nS_freq1}
\end{figure}

We performed comprehensive measurements of the absolute transition frequencies of the $5P_{3/2}, F'=3  \rightarrow nS$, $5P_{3/2}; F'=3 \rightarrow nD_{3/2}$ and $5P_{3/2}, F'=3 \rightarrow nD_{5/2}$ transitions to the Rydberg states of $^{87}Rb$ with principal quantum numbers $n = 45 -124$  for $nS_{1/2}$ and $n = 45-120 $ for $nD_{3/2}$ and $nD_{5/2}$ levels. The measured absolute frequencies for the  $5P_{3/2}, F'=3  \rightarrow nS$, $5P_{3/2}, F'=3 \rightarrow nD_{3/2}$ and $5P_{3/2},F'=3  \rightarrow nD_{5/2} $ transitions are tabulated in the Tables \ref{tab:nS1/2}, \ref{tab:nD3/2} and \ref{tab:nD5/2} respectively. The values in the parentheses for each of the transition frequencies listed in the tables are the overall uncertainties in the determination of the value of the corresponding transition frequency. For n$>$ 120, the SNR of the EIT signals decreases and the fine-structure splitting between the $nD_{3/2}$ and the $nD_{5/2}$ Rydberg levels becomes comparable to the EIT signal linewidths. Hence, we recorded the Ryberg EIT signals upto principal quantum number around n = 124.
\par
The variation of the fine structure splitting between the $nD_{3/2}$  and $nD_{5/2}$ Rydberg states with the principal quantum number n is shown in Fig.\ref{fig:nS_freq1}(a). The fine-structure splitting is expected to vary as $1/n^3$ with the principal quantum number n \cite{gallagher_1994, Greene_RMP}. Hence we fit the experimental data with $A/n^3$ with A as the fitting parameter as shown in Fig.\ref{fig:nS_freq1}(a). Fine structure splitting for three different principal quantum numbers n=46, n=55 and n=88 are shown in Fig.\ref{fig:nS_freq1}(b) depicting the decrease in the splitting for Rydberg states with larger values of n. 
\begin{figure}[h!]
\centering\includegraphics[scale=.2]{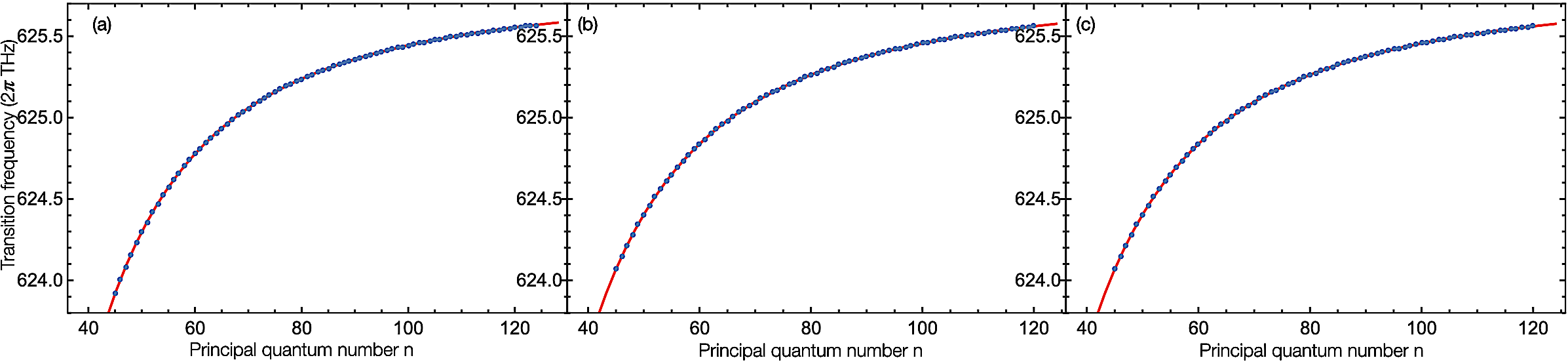}
\caption{Variation of the experimentally measured absolute transition frequency of the (a) $nS_{1/2}$ (b) $nD_{3/2}$  and (c)  $nD_{5/2}$ Rydberg states with the principal quantum number n. The error bars are smaller than the size of the data points. The transition frequencies plotted in (a) are the measured values of the transition frequencies of the $nS_{1/2}$ Rydberg states corrected for the hyperfine shift of the $nS_{1/2}$ Rydberg states using Eqn.{\ref{hfs4}}. The solid line is the fit to the data using Eq.\ref{eq:QD} which gives the values of the Rydberg-Ritz parameters tabulated in Table \ref{tab:QD}.}
\label{fig:nS_freq}
\end{figure}

\begin{table}[h!]
\label{tab:QD}
\centering \caption{The values of the quantum defects or the Rydberg Ritz parameters determined by fitting the data for the absolute transition frequencies for $nS_{1/2}$, $nD_{3/2}$ and $nD_{5/2}$ Rydberg levels with Eqn.\ref{eq:QD}. The measured values of the transition frequencies of the $nS_{1/2}$ Rydberg states were corrected for the hyperfine shift of the $nS_{1/2}$ Rydberg states using Eqn.{\ref{hfs4}} to estimate the values of the quantum defects.}
\begin{tabular*}{8cm}{c c c }
\hline \rule[-1ex]{0pt}{3.5ex} &\hspace{16 mm} $\delta_0$ &\hspace{16 mm} $\delta_2$\\
          \hline \rule[-1ex]{0pt}{3.5ex} $nS_{1/2}$ &\hspace{16 mm} 3.1311809(12) &\hspace{16 mm} 0.1786(6) \\ 
                    \rule[-1ex]{0pt}{3.5ex}  $nD_{3/2}$ &\hspace{16 mm} 1.3480942(15) &\hspace{16 mm} -0.6048(9) \\ 
                    \rule[-1ex]{0pt}{3.5ex}  $nD_{5/2}$&\hspace{16 mm} 1.3464628(14) &\hspace{16 mm} -0.5934(8)\\ 
                                      
\hline
\end{tabular*} 
\label{tab:QD}
\end{table}

\par
The variation of the absolute transition frequencies of the $5P_{3/2}, F'=3 \rightarrow nS$ and $5P_{3/2}, F'=3 \rightarrow nD_{3/2}$ and $5P_{3/2}, F'=3 \rightarrow nD_{5/2}$ transitions to the Rydberg states of $^{87}Rb$ with the principal quantum numbers n is shown in Fig. \ref{fig:nS_freq}(a), Fig. \ref{fig:nS_freq}(b) and Fig. \ref{fig:nS_freq}(c) respectively.
The values of the quantum defects or the Rydberg Ritz parameters for $^{87}$Rb were determined by fitting  the data for the absolute transition frequencies for $nS_{1/2}$, $nD_{3/2}$ and $nD_{5/2}$ Rydberg levels (as shown in Fig. \ref{fig:nS_freq}) with Eqn.{\ref{eq:QD} where the ionisation frequency ($E_i/\hbar$) was used as the common fitting parameter.  We subtract the hyperfine shift obtained from Eq.(7) from the measured transition frequencies of the $nS_{1/2}$ Rydberg states to find the corresponding Rydberg Ritz parameters. We tabulate the values of the Rydberg-Ritz parameters in Table \ref{tab:QD}. The values of the Rydberg-Ritz parameters for $^{87}$Rb obtained from our experimental data agree with those obtained in  \cite{2011_Mack} within our specified experimental uncertainty. The value of ionisation frequency of the $5P_{3/2}, F'=3$ state obtained from the fits to the experimental data for the transition frequencies is 625.7942146(5) $2\pi$ THz.

\begin{figure}[h!]
\centering\includegraphics[scale=.1]{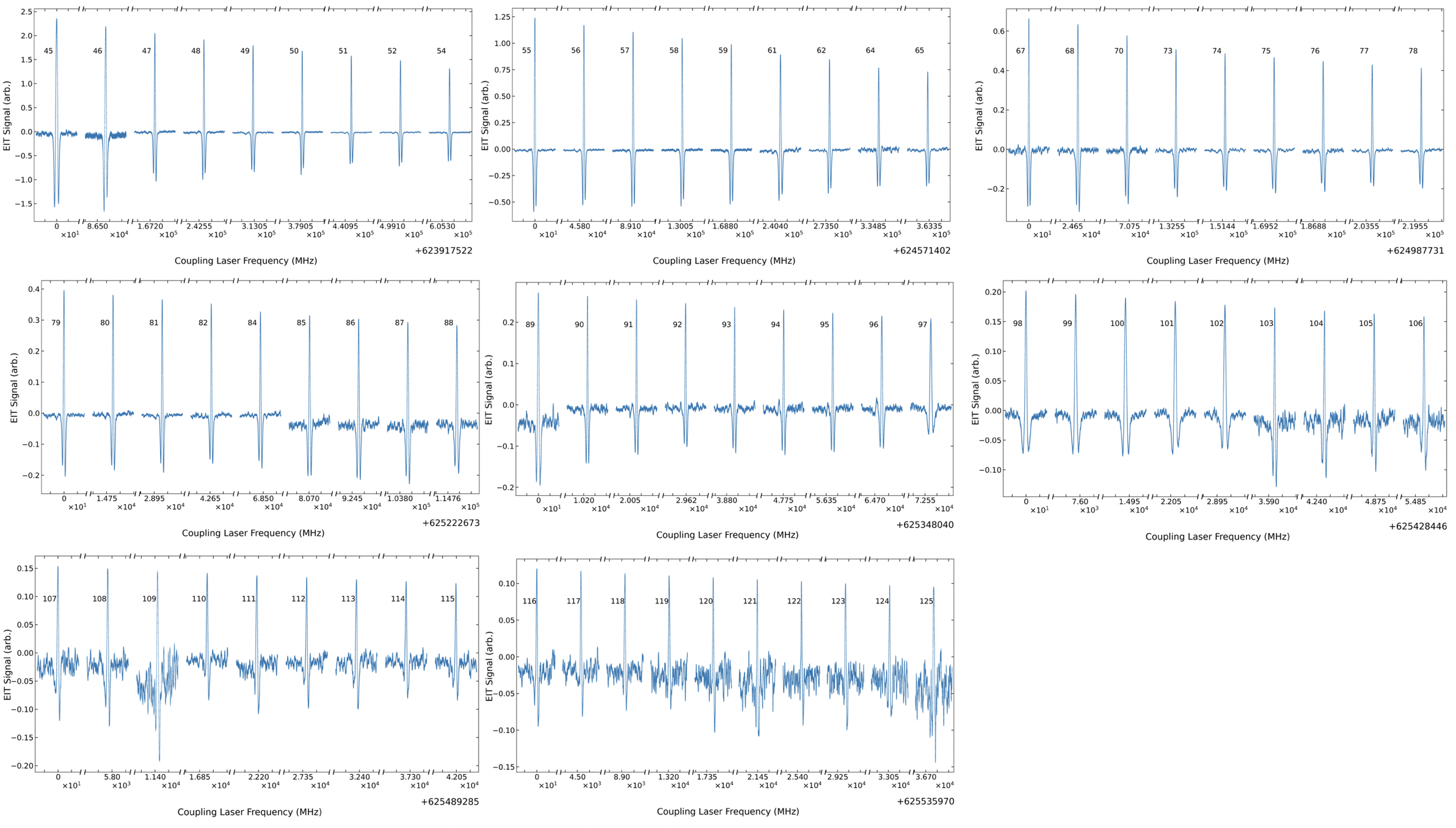}
\caption{Rydberg EIT data for transition to the $nS_{1/2}$ Rydberg states showing the variation of the transition strengths with the principal quantum number. All the signals shown were recorded using lock-in-detection via probe frequency modulation. We have scaled the Rydberg EIT signal amplitudes according to the gain factor used in the lock-in amplifier.}
\label{fig:nS_freq2}
\end{figure}

\begin{figure}[h!]
\centering\includegraphics[scale=.1]{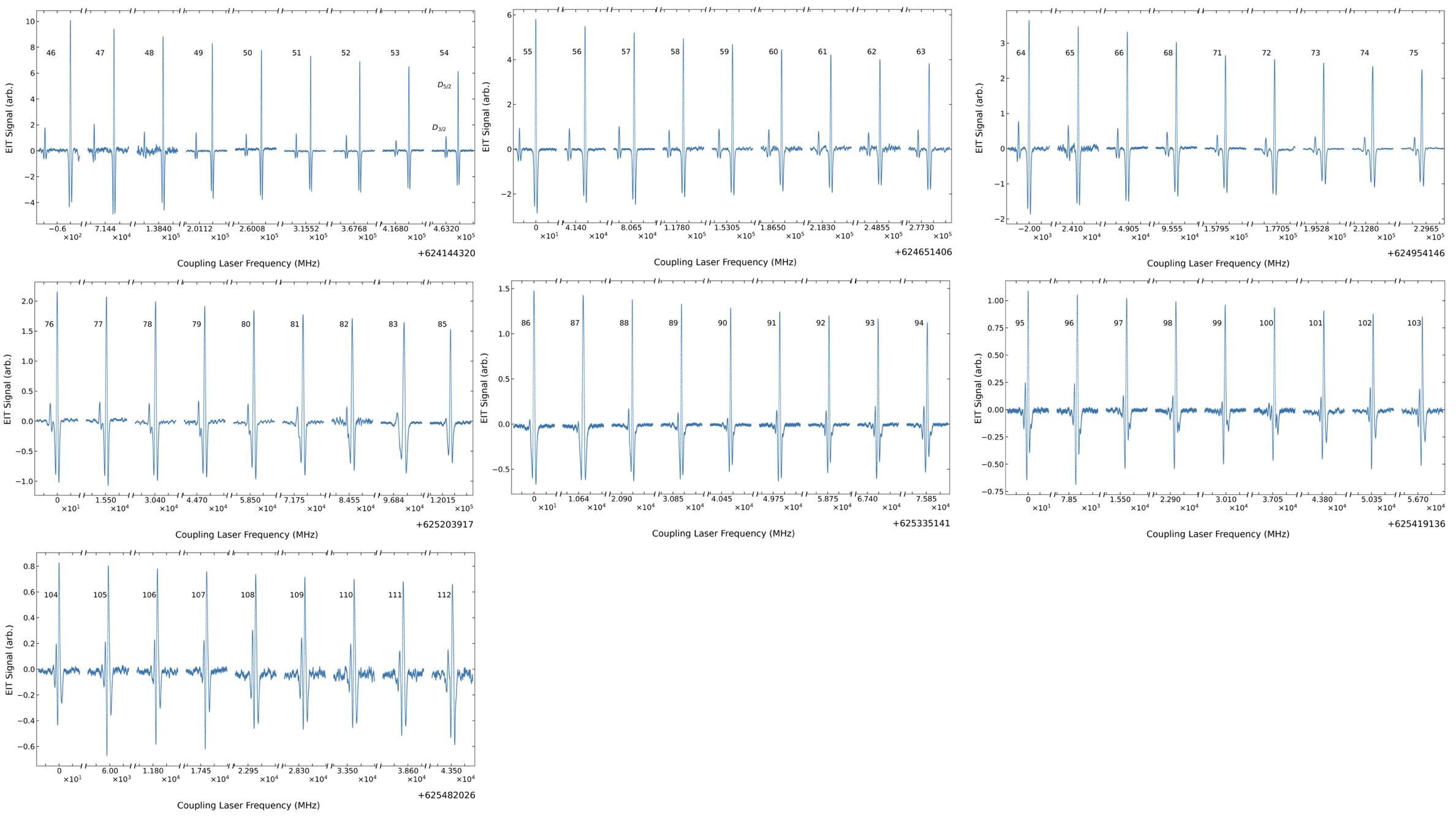}
\caption{Rydberg EIT data for transition to the $nD_{3/2}$ and $nD_{5/2}$ Rydberg states showing the variation of the transition strengths and fine-structure splitting with the principal quantum number. All the signals shown were recorded using lock-in-detection via probe frequency modulation. We have scaled the Rydberg EIT signal amplitudes according to the gain factor used in the lock-in amplifier.}
\label{fig:nD_freq3}
\end{figure}

\par
Rydberg EIT signals for transition to the Rydberg states depicting the variation of the transition strengths with the principal quantum number is shown in Fig.\ref{fig:nS_freq2} for $nS_{1/2}$ Rydberg states and in Fig.\ref{fig:nD_freq3} for $nD_{3/2}$ and $nD_{5/2}$ Rydberg states. The experimental data in Fig.\ref{fig:nD_freq3} also shows the variation of the fine-structure splitting between the $nD_{3/2}$  and $nD_{5/2}$ Rydberg states with the values of n. The data shown in Fig.\ref{fig:nS_freq2} and Fig.\ref{fig:nD_freq3} were all recorded using lock-in detection via frequency modulation of the probe beam. We have scaled the Rydberg EIT signal amplitudes according to the gain factor used in the lock-in amplifier. The variation of the transition strengths of the transition to the Ryberg states with the principal quantum number is expected to vary as $1/n^{*3}$ \cite{gallagher_1994, Greene_RMP} where $n^*=n-\delta_{nlj}$ is the effective quantum number. From our experimental data, we observe a similar qualitative behavior with transition strengths decreasing with increasing effective principal quantum number as shown in Fig.\ref{fig:nS_freq2} and Fig.\ref{fig:nD_freq3}.


\section{Conclusion and Outlook} \label{sec:Conclusion_Outlook}
In this work, we perform a comprehensive measurement of the absolute transition frequency to nS and nD Rydberg states of $^{87}$Rb with high principal quantum numbers n=45-124 in a wide range of values. We determine the Rydberg-Ritz parameters for $^{87}$Rb from the fits of the data for the measured transition frequencies for each of the data set corresponding to the transitions to $nS_{1/2}$, $nD_{3/2}$ and $nD_{5/2}$ Rydberg levels. We also obtain the value of the ionisation frequency for $5P_{3/2}, F=3$ state of $^{87}Rb$ as the fitting parameter for the fits. 
\par
Our measurements provide a framework for Rydberg excitation to highly excited Rydberg states in cold atom ensembles\cite{2021_Weidemuller} as well as in arrays of Rydberg atom trapped in optical tweezers\cite{2021_Lukin}. Such cold atom systems with long-lived coherence are ideal toolbox for quantum sensors as we demonstrated recently in the context of spin coherence in cold atoms\cite{2021_PRR}. Using highly excited cold Rydberg atoms with large polarizability ($\sim n^7$), precision electrometry\cite{2017_electrometry} can be performed analogous to precision magnetometry\cite{2018_OE} using spin correlation spectroscopy. Our measurements of the absolute transition frequencies of the highly excited nS and nD Rydberg states of $^{87}Rb$ would be useful for the identification of these states for atomic physics experiments as well as quantum computing\cite{Saffman_2016} and quantum simulation\cite{2021_Browaeys} with Rydberg atoms with exaggerated values of the long-range dipole-dipole interaction. 

\section*{Appendix}
\begin{appendix} 
\renewcommand{\thesection}{\Alph{section}}

\section{Estimation of Hyperfine splitting of $nS_{1/2}$ Rydberg states}
\label{sec_hfs}
The hyperfine shift in terms of the quantum numbers I, J and F for the nuclear spin, electronic angular momentum and total angular momentum respectively is given by:
\begin{equation}
\Delta_{hfs, F} = \frac{A}{2} [F(F+1) - I(I+1) -J(J+1)] 
\label{hfs2}
\end{equation}
where A is the hyperfine constant. For the $nS_{1/2}$ Rydberg states of $^{87}Rb $, J=1/2, I = 3/2 and hence, the hyperfine splitting between the levels $F= I+1/2$ and $F=I-1/2$ is given by:
\begin{equation}
\Delta E_{hf}= \Delta_{hfs, F=2} - \Delta_{hfs, F=1} =(I+\frac{1}{2})A=2A
\label{hfs1}
\end{equation}
Using Eqn.{\ref{hfs}} and Eqn.{\ref{hfs1}}, we get 
\begin{equation}
2A=37.1(2)\, \text{GHz} \,  \times {n^{*}}^{-3}
\label{hfs3}
\end{equation}

\section{Measurement of frequency uncertainty of probe laser}
\label{freq_stability}
 To measure the frequency uncertainty of the probe laser, the laser was locked to the ($5S_{1/2}, F=2  \rightarrow 5P_{3/2}, F'=3$ transition of $^{87}$Rb. The laser frequency was measured using the wavelength meter at repeated intervals after locking the laser. The histogram of the measured frequency is shown in Fig.{\ref{probe_frequency}}. A frequency uncertainty of $\approx$ 400 kHz of the probe laser was obtained from the FWHM obtained from the Gaussian fit to the data.
 
  \begin{figure}[hbtp]
    \centering
    \includegraphics[scale=.2]{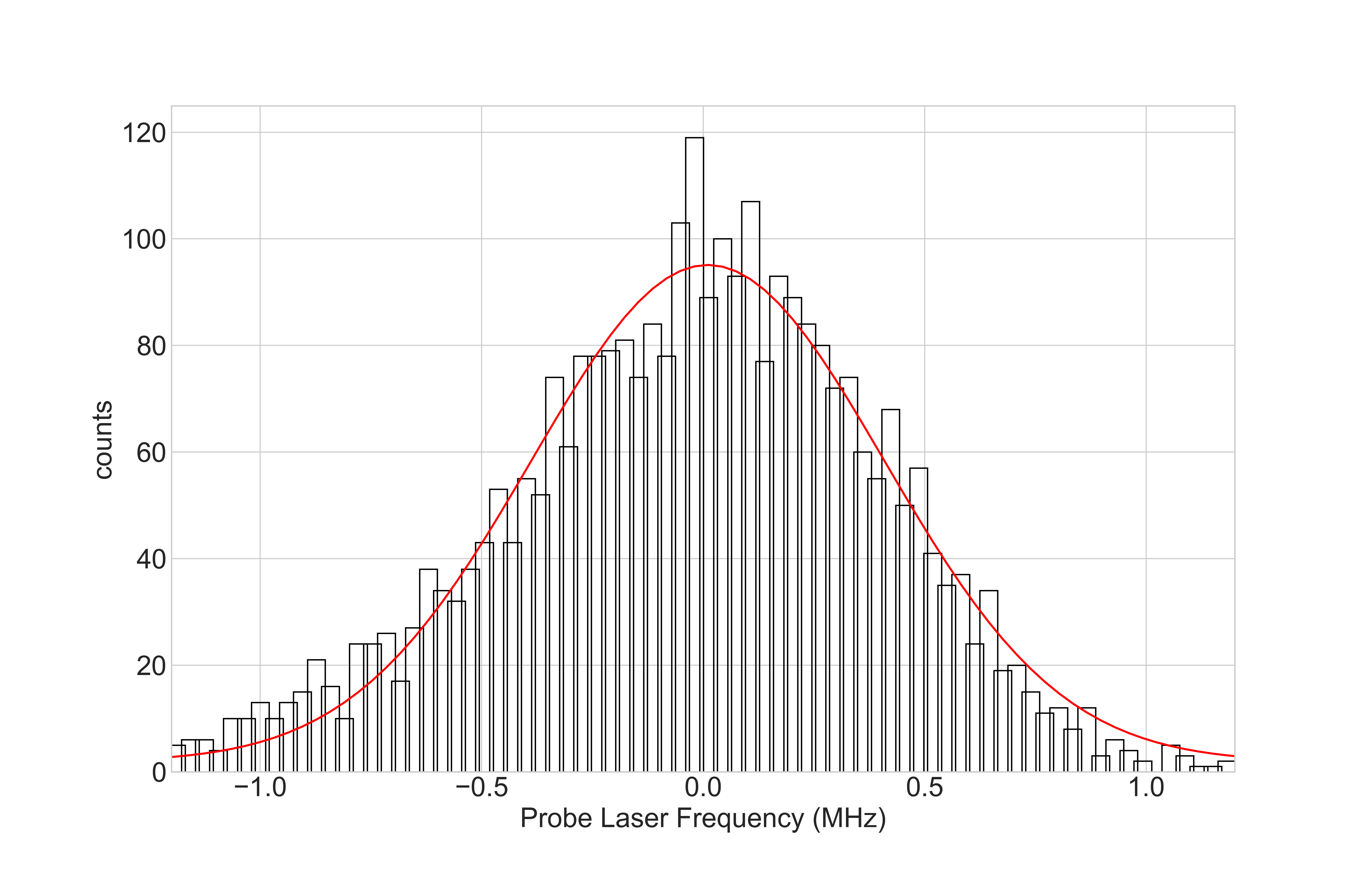}
    \caption{Histogram of the measured frequency of the probe laser after locking to the crossover between the ($5S_{1/2} (F=2) \rightarrow 5P_{3/2}(F'=2)$ transition and the ($5S_{1/2} (F=2)  \rightarrow 5P_{3/2}(F'=3)$ transition of $^{87}$Rb. The FWHM obtained from the Gaussian fit to the data gives a frequency uncertainty $\approx$ 400 kHz for the probe laser.}
     \label{probe_frequency}
    \end{figure}

 \begin{figure}[hbtp]
    \centering
    \includegraphics[scale=.2 ]{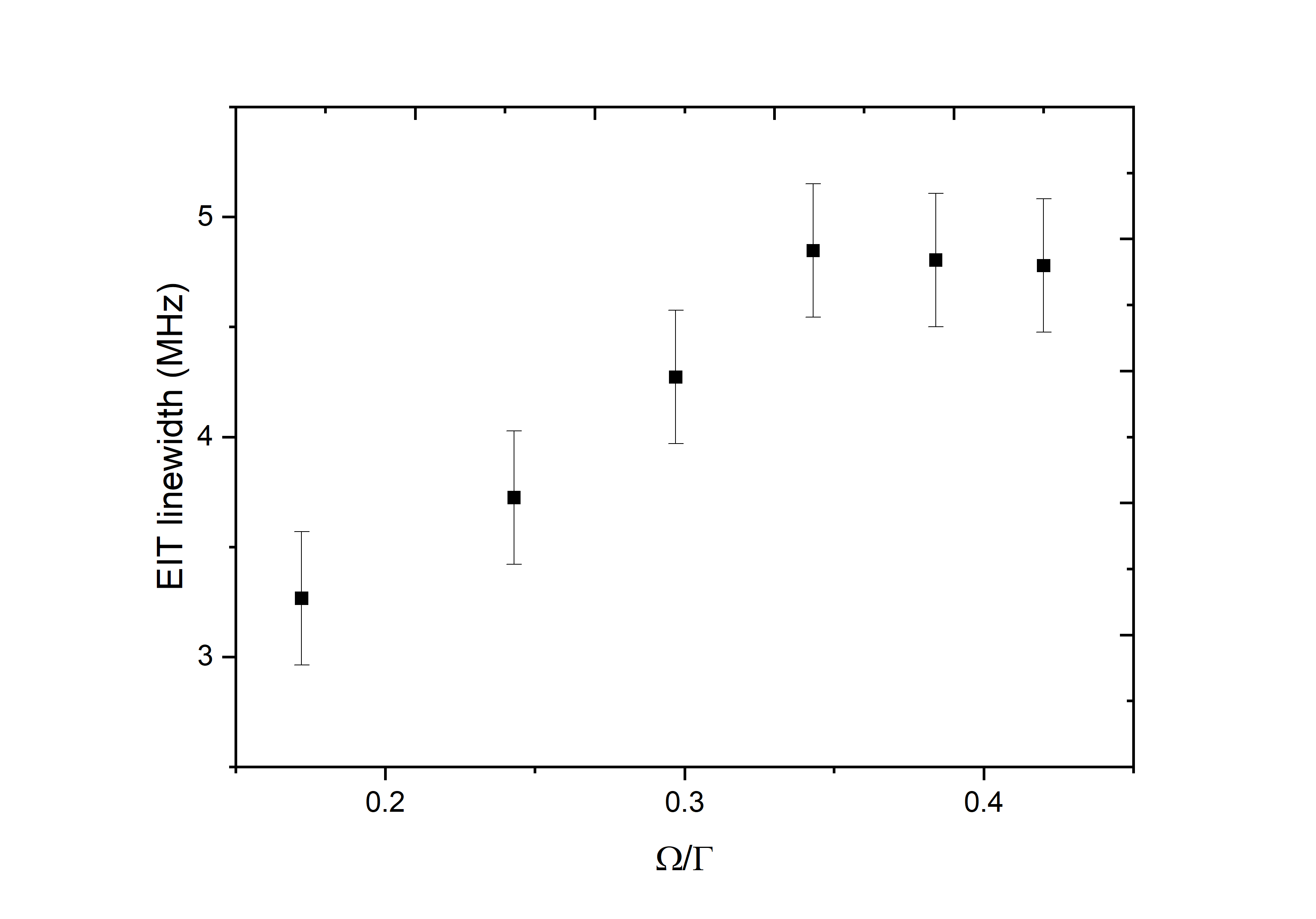}
    \caption{Variation of the EIT signal linewidth as a function of the normalised Rabi frequency ($\Omega/\Gamma$) of the probe beam with the coupling beam power fixed at 8 mW.}
     \label{EIT_width_probe_power}
    \end{figure}

\section{Fitting of Lock-in detection signal}
\label{sec_fit}
We have used lock-in detection for recording the Rydberg EIT signals for n$>$80 to maintain a good signal to noise ratio.  Due to the asymmetricity of the signal recorded using lock-in detection, the second derivative of symmetry functions like Gaussian and Lorentzian does not properly fit the signals recorded using lock-in detection. Hence, we have used the second derivative of the asymmetric pseudo-Voigt function to fit the Rydberg EIT signals recorded via lock-in detection to accurately determine the transition frequencies to the Rydberg states.	

\begin{figure}[hbtp]
 \centering
 \includegraphics[scale=.4]{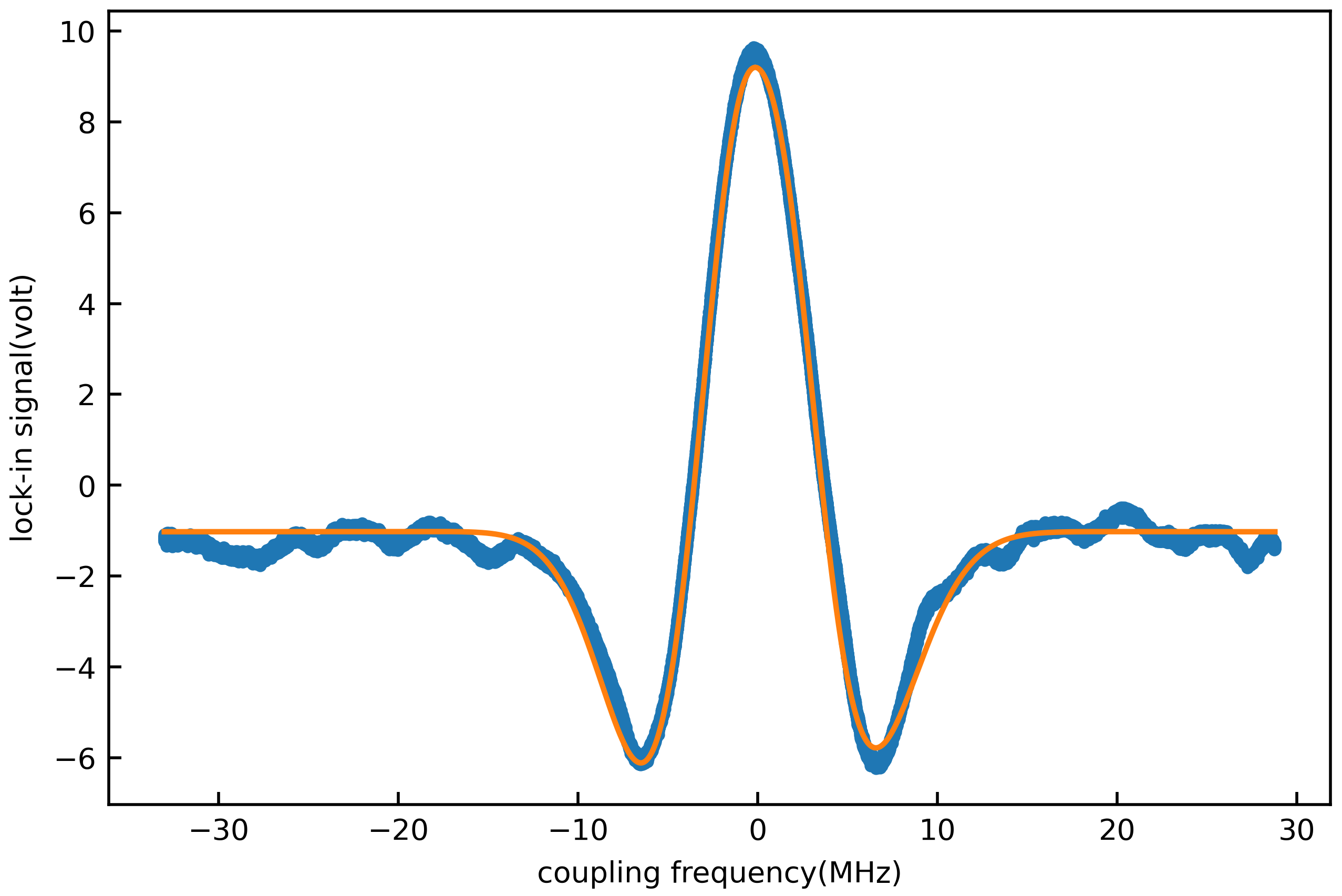}
 \caption{A typical probe transmission signal at the Rydberg EIT resonance for $88S_{1/2}$ Rydberg state measured using lock-in detection with the coupling laser frequency being scanned across the EIT resonance. The  signal was fitted with a pseudo-Voigt function (solid line)  to obtain the coupling laser frequencies corresponding to the peak of the signal corresponding to the Rydberg states.}
 \label{coupling_scan}
 \end{figure}

\par
A true Voigt function does not have a closed form, so taking differentiation and fitting will be impractical. Instead, we used a pseudo-Voigt function which is basically a linear combination of Lorentzian and Gaussian profile given by: \\
\begin{equation}
Y(x,x_0)= \mu L(x,x_0) +(1-\mu) G(x,x_0)
\end{equation}
 where, $\mu$ is the fractional contribution of the Lorentzian lineshape; $L(x,x_0)$ and $G(x,x_0)$ are the Lorentzian and Gaussian profiles respectively peaked at $x_0$ given by:
 \begin{equation}
L(x,x_0)=\frac{A_1}{(\frac{4(x-x_0^2)^2}{\gamma_0 ^2}+1)}
\end{equation}
\begin{equation}
G(x,x_0)=A_2 e^{-4log2\frac{(x-x_0)^2}{\gamma_0 ^2}}
\end{equation}
where, $\gamma_0$ is the linewidth of the profiles.
\par
A sigmoid function was used to bring the asymmetry in both the lineshape functions:
\begin{equation}
\gamma(x,x_0)=\frac{2\gamma_0}{1+e^{a(x-x_0)}}
\end{equation}
where a is the asymmetry parameter giving the measure of the asymmetry. $\gamma(x,x_0)$ was substituted in place of  $\gamma_0$ in the lineshape functions to form an asymmetric pseudo-Voigt function. 
The second-order derivative of this asymmetric pseudo-Voigt function was fitted with the Rydberg EIT signals recorded using lock-in detection to determine the peak of the Rydberg EIT signal and thereby the absolute transition frequencies to the Rydberg levels accurately.
    
\end{appendix}
\begin{table}[t]
\centering \caption{The experimentally measured absolute transition frequencies of the  $5P_{3/2} (F'=3) \rightarrow nS$ transitions.}
\scalebox{0.88}
{
\begin{tabular*}{10cm}{c c c c c }
\hline \rule[-1ex]{0pt}{2ex} n &\hspace{8 mm} $\omega_{expt}$ \text{($2\pi$ THz)}& \vline &  \hspace{4mm} n &\hspace{8 mm} $\omega_{expt}$ \text{($2\pi$ THz)}\\
          \hline \rule[-1ex]{0pt}{1ex} 45 &\hspace{8 mm} 623.9175224(16) &  \vline & \hspace{4mm}  85 &\hspace{8mm} 625.3033775(18) \\ 
                    \rule[-1ex]{0pt}{2ex} 46 &\hspace{8 mm} 624.0040574(12) & \vline & \hspace{4mm} 86 &\hspace{8mm} 625.3151519(19) \\ 
                    \rule[-1ex]{0pt}{2ex} 47 &\hspace{8 mm} 624.0847416(10) & \vline & \hspace{4mm} 87 &\hspace{8mm} 625.3265079(19) \\ 
                    \rule[-1ex]{0pt}{2ex} 48 &\hspace{8 mm} 624.1600918(09)  & \vline & \hspace{4mm} 88 &\hspace{8mm} 625.3374648(18)\\                   
                    \rule[-1ex]{0pt}{3.5ex} 49 &\hspace{8 mm} 624.2305679(12) &  \vline &\hspace{4mm} 89 &\hspace{8mm} 625.3480410(19) \\ 
                    \rule[-1ex]{0pt}{3.5ex} 50 &\hspace{8 mm} 624.2965805(09)  & \vline &\hspace{4mm}  90 &\hspace{8mm} 625.3582558(19) \\ 
                    \rule[-1ex]{0pt}{3.5ex} 51 &\hspace{8 mm} 624.3584991(11) & \vline &\hspace{4mm} 91 &\hspace{8mm} 625.3681210(18) \\ 
                    \rule[-1ex]{0pt}{3.5ex} 52 &\hspace{8 mm} 624.4166576(15) & \vline &\hspace{4mm} 92 &\hspace{8mm} 625.3776596(20)\\ 
                    \rule[-1ex]{0pt}{3.5ex} 53 &\hspace{8 mm} 624.4713495 (11) & \vline &\hspace{4mm} 93 &\hspace{8mm} 625.3868760(15) \\ 
                    \rule[-1ex]{0pt}{3.5ex} 54 &\hspace{8 mm} 624.5228507(18) & \vline  &\hspace{4mm} 94&\hspace{8mm} 625.3957927(17) \\
                    \rule[-1ex]{0pt}{3.5ex} 55 &\hspace{8 mm} 624.5714018(19) & \vline   &\hspace{4mm} 95 &\hspace{8mm} 625.4044195(12) \\
                    \rule[-1ex]{0pt}{3.5ex} 56 &\hspace{8 mm} 624.6172213(12) & \vline   &\hspace{4mm} 96 &\hspace{8mm} 625.4127680(14)\\
                    \rule[-1ex]{0pt}{3.5ex} 57 &\hspace{8 mm} 624.6605137(09) & \vline  &\hspace{4mm} 97 &\hspace{8mm} 625.4208520(11)\\
                    \rule[-1ex]{0pt}{3.5ex} 58 &\hspace{8 mm} 624.7014609(14) & \vline  &\hspace{4mm} 98 &\hspace{8mm} 625.4286830(15)\\
                    \rule[-1ex]{0pt}{3.5ex} 59 &\hspace{8 mm} 624.7402289(16) & \vline  &\hspace{4mm} 99 &\hspace{8mm} 625.4362696(19)\\
                    \rule[-1ex]{0pt}{3.5ex} 60 &\hspace{8 mm} 624.7769710(12) & \vline &\hspace{4mm} 100&\hspace{8mm} 625.4436187(16)\\
                    \rule[-1ex]{0pt}{3.5ex} 61 &\hspace{8 mm} 624.8118251(18) & \vline &\hspace{4mm} 101&\hspace{8mm} 625.4507467(17) \\
                    \rule[-1ex]{0pt}{3.5ex} 62 &\hspace{8 mm} 624.8449181(19) & \vline &\hspace{4mm} 102&\hspace{8mm} 625.4576591(18)\\
                    \rule[-1ex]{0pt}{3.5ex} 63 &\hspace{8 mm} 624.8763642(13) & \vline &\hspace{4mm} 103&\hspace{8mm} 625.4643650(19)\\
                    \rule[-1ex]{0pt}{3.5ex} 64 &\hspace{8 mm} 624.9062754(10) & \vline &\hspace{4mm} 104&\hspace{8mm} 625.4708747(18)\\
                    \rule[-1ex]{0pt}{3.5ex} 65 &\hspace{8 mm} 624.9347472(13) & \vline &\hspace{4mm} 105&\hspace{8mm} 625.4771918(17)\\
                    \rule[-1ex]{0pt}{3.5ex} 66 &\hspace{8 mm} 624.9618711(11) & \vline &\hspace{4mm} 106&\hspace{8mm} 625.4833270(19) \\
                    \rule[-1ex]{0pt}{3.5ex} 67 &\hspace{8 mm} 624.9877317(10) & \vline &\hspace{4mm}107&\hspace{8mm} 625.4892832(18)\\
                    \rule[-1ex]{0pt}{3.5ex} 68 &\hspace{8 mm} 625.0124047(09) & \vline &\hspace{4mm}108&\hspace{8mm}  625.4950696(16)\\
                    \rule[-1ex]{0pt}{3.5ex} 69 &\hspace{8 mm} 625.0359629(12) & \vline &\hspace{4mm}109&\hspace{8mm}  625.5006945(18)\\
                    \rule[-1ex]{0pt}{3.5ex} 70 &\hspace{8 mm} 625.0584728(14) & \vline &\hspace{4mm}110&\hspace{8mm}  625.5061623(15) \\
                    \rule[-1ex]{0pt}{3.5ex} 71 &\hspace{8 mm} 625.0799937(11) & \vline &\hspace{4mm}111&\hspace{8mm}  625.5114789(17)\\
                    \rule[-1ex]{0pt}{3.5ex} 72 &\hspace{8 mm} 625.1005846(12) & \vline &\hspace{4mm}112&\hspace{8mm} 625.5166491(15) \\
                    \rule[-1ex]{0pt}{3.5ex} 73 &\hspace{8 mm} 625.1202986(14) &  \vline&\hspace{4mm}113&\hspace{8mm} 625.5216765(19)\\                             
                    \rule[-1ex]{0pt}{3.5ex} 74 &\hspace{8 mm} 625.1391807(17) & \vline &\hspace{4mm}114&\hspace{8mm} 625.5265710(17)\\
                    \rule[-1ex]{0pt}{3.5ex}75 &\hspace{8 mm} 625.1572829(15) & \vline  &\hspace{4mm}  115&\hspace{8mm} 625.5313353(18)\\                          
                    \rule[-1ex]{0pt}{3.5ex}76 &\hspace{8 mm} 625.1746455(14) & \vline &\hspace{4mm}  116 &\hspace{8mm} 625.5359720(19)\\ 
                    \rule[-1ex]{0pt}{3.5ex}77 &\hspace{8 mm} 625.1913050(19) & \vline &\hspace{4mm}  117 &\hspace{8mm} 625.5404877(18)\\
                    \rule[-1ex]{0pt}{3.5ex} 78 &\hspace{8 mm}625.2073056(11) & \vline &\hspace{4mm}118 &\hspace{8mm} 625.5448867(19)\\
                    \rule[-1ex]{0pt}{3.5ex} 79 &\hspace{8 mm}625.2226738(12) & \vline &\hspace{4mm} 119 &\hspace{8mm} 625.5491732(19)\\
                    \rule[-1ex]{0pt}{3.5ex} 80 &\hspace{8 mm}625.2374478(15) & \vline &\hspace{4mm} 120 &\hspace{8mm} 625.5533480(18)\\
                    \rule[-1ex]{0pt}{3.5ex} 81 &\hspace{8 mm}625.2516554(19)&  \vline &\hspace{4mm}121&\hspace{8mm} 625.5574184(17)\\
                    \rule[-1ex]{0pt}{3.5ex} 82 &\hspace{8 mm}625.2653287(15)&\vline  &\hspace{4mm} 122 &\hspace{8mm} 625.5613862(19)\\
                    \rule[-1ex]{0pt}{3.5ex}83 &\hspace{8 mm} 625.2784893(13)&\vline&\hspace{4mm}123&\hspace{8mm} 625.5652550(16)\\
                    \rule[-1ex]{0pt}{3.5ex}84&\hspace{8 mm}625.2911636(17)&\vline&\hspace{4mm}124&\hspace{8mm}625.5690249(18)\\
 \hline
\end{tabular*} 
}
\label{tab:nS1/2}
\end{table}

\begin{table}[t]

\centering \caption{The experimentally measured absolute transition frequencies of the  $5P_{3/2} (F'=3) \rightarrow nD_{3/2}$ transitions.}
\scalebox{0.88}
{
\begin{tabular*}{10cm}{c c c c c }
\hline \rule[-1ex]{0pt}{1ex} n &\hspace{8 mm}$\omega_{expt}$ \text{($2\pi$ THz)}& \vline &  \hspace{4mm} n &\hspace{8 mm} $\omega_{expt}$ \text{($2\pi$ THz)}\\                                                                                                              
         \hline \rule[-1ex]{0pt}{3.5ex} 45 &\hspace{8 mm} 624.0677419(10)  &  \vline & \hspace{4mm}  83&\hspace{8mm} 625.3007683(19) \\
                     \rule[-1ex]{0pt}{3.5ex} 46 &\hspace{8 mm} 624.1442057(12)&  \vline & \hspace{4mm}84 &\hspace{8mm}625.3126371(15)   \\ 
                    \rule[-1ex]{0pt}{3.5ex} 47 &\hspace{8 mm} 624.2156989(19) & \vline & \hspace{4mm}  85&\hspace{8mm} 625.3240816(19) \\ 
                    \rule[-1ex]{0pt}{3.5ex} 48 &\hspace{8 mm} 624.2826429(13)  & \vline & \hspace{4mm} 86&\hspace{8mm}625.3351245(17) \\ 
                    \rule[-1ex]{0pt}{3.5ex} 49 &\hspace{8 mm} 624.3454194(18) & \vline & \hspace{4mm} 87&\hspace{8mm} 625.3457814(18) \\                   
                    \rule[-1ex]{0pt}{3.5ex} 50 &\hspace{8 mm} 624.4043647(19) &  \vline &\hspace{4mm} 88 &\hspace{8mm}  625.3560721(16) \\ 
                    \rule[-1ex]{0pt}{3.5ex} 51 &\hspace{8 mm} 624.4597838(17)  & \vline &\hspace{4mm}89 &\hspace{8mm} 625.3660134(19)  \\ 
                    \rule[-1ex]{0pt}{3.5ex} 52 &\hspace{8 mm} 624.5119537(14) & \vline &\hspace{4mm} 90&\hspace{8mm} 625.3756175(18) \\ 
                     \rule[-1ex]{0pt}{3.5ex} 53 &\hspace{8 mm} 624.5611229(19) & \vline &\hspace{4mm} 91&\hspace{8mm} 625.3849041(17) \\ 
                    \rule[-1ex]{0pt}{3.5ex} 54 &\hspace{8 mm} 624.6075171(19) & \vline &\hspace{4mm} 92&\hspace{8mm}  625.3938854(18) \\ 
                    \rule[-1ex]{0pt}{3.5ex} 55 &\hspace{8 mm} 624.6513393(18) & \vline  &\hspace{4mm}93 &\hspace{8mm} 625.4025737(19)\\
                    \rule[-1ex]{0pt}{3.5ex} 56 &\hspace{8 mm} 624.6927805(16)  & \vline   &\hspace{4mm}94&\hspace{8mm} 625.4109827(19) \\
                    \rule[-1ex]{0pt}{3.5ex} 57 &\hspace{8 mm} 624.7320072(14) & \vline   &\hspace{4mm}95 &\hspace{8mm} 625.4191226(17) \\
                     \rule[-1ex]{0pt}{3.5ex} 58 &\hspace{8 mm} 624.7691757(12) & \vline  &\hspace{4mm}  96&\hspace{8mm} 625.4270060(19) \\
                     \rule[-1ex]{0pt}{3.5ex} 59 &\hspace{8 mm} 624.8044251(15) & \vline  &\hspace{4mm}97&\hspace{8mm}625.4346435(18) \\
                    \rule[-1ex]{0pt}{3.5ex} 60 &\hspace{8 mm} 624.8378902(14)  & \vline  &\hspace{4mm} 98&\hspace{8mm} 625.4420457(16) \\
                    \rule[-1ex]{0pt}{3.5ex} 61 &\hspace{8 mm} 624.8696863(19)& \vline &\hspace{4mm}  99&\hspace{8mm} 625.4492206(15) \\
                    \rule[-1ex]{0pt}{3.5ex} 62 &\hspace{8 mm} 624.8999202(18)& \vline &\hspace{4mm} 100&\hspace{8mm} 625.4561797(19) \\
                    \rule[-1ex]{0pt}{3.5ex} 63 &\hspace{8 mm} 624.9286928(16) & \vline &\hspace{4mm} 101&\hspace{8mm}625.4629307(17) \\
                    \rule[-1ex]{0pt}{3.5ex} 64 &\hspace{8 mm} 624.9561033(15) & \vline &\hspace{4mm} 102&\hspace{8mm} 625.4694809(18) \\
                    \rule[-1ex]{0pt}{3.5ex} 65 &\hspace{8 mm} 624.9822305(12) & \vline &\hspace{4mm}103&\hspace{8mm} 625.4758392(18)\\
                   \rule[-1ex]{0pt}{3.5ex} 66 &\hspace{8 mm} 625.0071536(17) & \vline &\hspace{4mm} 104&\hspace{8mm} 625.4820116(19) \\
                   \rule[-1ex]{0pt}{3.5ex} 67 &\hspace{8 mm} 625.0309483(14) & \vline &\hspace{4mm}  105&\hspace{8mm} 625.4880061(15)  \\
                   \rule[-1ex]{0pt}{3.5ex} 68 &\hspace{8 mm} 625.0536785(16)& \vline &\hspace{4mm} 106&\hspace{8mm} 625.4938297(14) \\
                   \rule[-1ex]{0pt}{3.5ex} 69 &\hspace{8 mm} 625.0754099(13) & \vline &\hspace{4mm}107&\hspace{8mm} 625.4994907(19)\\
                   \rule[-1ex]{0pt}{3.5ex} 70 &\hspace{8 mm} 625.0961978(15) & \vline &\hspace{4mm}108&\hspace{8mm} 625.5049894(16)\\
                   \rule[-1ex]{0pt}{3.5ex} 71 &\hspace{8 mm} 625.1160981(19) & \vline &\hspace{4mm}109&\hspace{8mm} 625.5103382(18)\\
                   \rule[-1ex]{0pt}{3.5ex} 72 &\hspace{8 mm} 625.1351570(19) & \vline &\hspace{4mm}110&\hspace{8mm}625.5155395(19) \\
                   \rule[-1ex]{0pt}{3.5ex} 73 &\hspace{8 mm} 625.1534231(17) & \vline &\hspace{4mm}111 &\hspace{8mm}625.5205981(15) \\
                   \rule[-1ex]{0pt}{3.5ex} 74 &\hspace{8 mm} 625.1709429(15) &  \vline &\hspace{4mm}112&\hspace{8mm} 625.5255208(18)\\
                   \rule[-1ex]{0pt}{3.5ex} 75 &\hspace{8 mm} 625.1877544(20) & \vline  &\hspace{4mm} 113&\hspace{8mm} 625.5303126(19) \\
                   \rule[-1ex]{0pt}{3.5ex} 76 &\hspace{8 mm} 625.2038910(18) & \vline &\hspace{4mm}114&\hspace{8mm} 625.5349765(17) \\
                   \rule[-1ex]{0pt}{3.5ex} 77&\hspace{8 mm} 625.2193954(14)& \vline &\hspace{4mm} 115&\hspace{8mm} 625.5395188(16)\\
                   \rule[-1ex]{0pt}{3.5ex} 78 &\hspace{8 mm} 625.2342960(17) & \vline &\hspace{4mm}116&\hspace{8mm}625.5439421(18) \\
                   \rule[-1ex]{0pt}{3.5ex} 79 &\hspace{8 mm} 625.2486236(15) & \vline &\hspace{4mm}117&\hspace{8mm} 625.5482507(19)\\
                   \rule[-1ex]{0pt}{3.5ex} 80 &\hspace{8 mm} 625.2624091(14) & \vline&\hspace{4mm} 118&\hspace{8mm} 625.5524503(18) \\
                    \rule[-1ex]{0pt}{3.5ex} 81&\hspace{8 mm} 625.2756795(19) & \vline&\hspace{4mm}119&\hspace{8mm} 625.5565425(20) \\
                   \rule[-1ex]{0pt}{3.5ex} 82&\hspace{8 mm} 625.2884578(18) & \vline &\hspace{4mm} 120&\hspace{8mm}625.5605318(19)\\

\hline 
\end{tabular*} 
}
\label{tab:nD3/2}
\end{table}

\begin{table}[t]

\centering \caption{The experimentally measured absolute transition frequencies of the  $5P_{3/2} (F'=3) \rightarrow nD_{5/2}$ transitions.}
\scalebox{0.88}
{
\begin{tabular*}{10cm}{c c c c c }
\hline \rule[-3ex]{0pt}{3.5ex} n &\hspace{8 mm} $\omega_{expt}$ \text{($2\pi$ THz)}& \vline &  \hspace{4mm} n &\hspace{8 mm} $\omega_{expt}$ \text{($2\pi$ THz)}\\
                                                                                                                   
           \hline \rule[-1ex]{0pt}{3.5ex} 45 &\hspace{8 mm} 624.0678706(12) &\vline& \hspace{4mm} 83&\hspace{8mm} 625.3007873(17) \\
                     \rule[-1ex]{0pt}{3.5ex} 46 &\hspace{8 mm} 624.1443250(19)&\vline & \hspace{4mm} 84&\hspace{8mm} 625.3126561(16) \\ 
                    \rule[-1ex]{0pt}{3.5ex} 47 &\hspace{8 mm} 624.2158095(14) &\vline & \hspace{4mm} 85&\hspace{8mm}625.3241004(19)\\ 
                    \rule[-1ex]{0pt}{3.5ex} 48 &\hspace{8 mm} 624.2827490(19) & \vline & \hspace{4mm}86&\hspace{8mm} 625.3351416(17)\\ 
                    \rule[-1ex]{0pt}{3.5ex} 49&\hspace{8 mm} 624.3455185(13)  & \vline & \hspace{4mm}87 &\hspace{8mm} 625.3457978(16)\\                   
                    \rule[-1ex]{0pt}{3.5ex} 50&\hspace{8 mm} 624.4044567(15)  &  \vline &\hspace{4mm}88&\hspace{8mm} 625.3560895(18) \\ 
                    \rule[-1ex]{0pt}{3.5ex} 51&\hspace{8 mm} 624.4598690(18)  & \vline &\hspace{4mm} 89&\hspace{8mm} 625.3660291(19)\\ 
                    \rule[-1ex]{0pt}{3.5ex} 52&\hspace{8 mm} 624.5120357(16) & \vline &\hspace{4mm} 90&\hspace{8mm} 625.3756314(18)\\ 
                     \rule[-1ex]{0pt}{3.5ex}53&\hspace{8 mm} 624.5611997(12) & \vline &\hspace{4mm} 91&\hspace{8mm} 625.3849202(17) \\ 
                    \rule[-1ex]{0pt}{3.5ex} 54&\hspace{8 mm} 624.6075896(19) & \vline &\hspace{4mm} 92&\hspace{8mm} 625.3938979(17)\\ 
                    \rule[-1ex]{0pt}{3.5ex} 55&\hspace{8 mm} 624.6514092(19) & \vline &\hspace{4mm}93&\hspace{8mm} 625.4025878(15)\\
                    \rule[-1ex]{0pt}{3.5ex} 56 &\hspace{8 mm}624.6928443(17) & \vline &\hspace{4mm}94&\hspace{8mm} 625.4109948(14)\\
                    \rule[-1ex]{0pt}{3.5ex}  57&\hspace{8 mm} 624.7320685(19)& \vline &\hspace{4mm}95&\hspace{8mm}625.4191350(18)\\
                     \rule[-1ex]{0pt}{3.5ex} 58 &\hspace{8 mm}624.7692333(12)& \vline &\hspace{4mm}96&\hspace{8mm} 625.4270198(19)\\
                     \rule[-1ex]{0pt}{3.5ex} 59 &\hspace{8 mm}624.8044814(18)& \vline &\hspace{4mm} 97&\hspace{8mm} 625.4346560(19)\\
                    \rule[-1ex]{0pt}{3.5ex} 60 &\hspace{8 mm} 624.8379430(12)& \vline &\hspace{4mm} 98 &\hspace{8mm}625.4420579(18)\\
                    \rule[-1ex]{0pt}{3.5ex} 61 &\hspace{8 mm} 624.8697366(19)& \vline &\hspace{4mm} 99&\hspace{8mm} 625.4492336(20)\\
                    \rule[-1ex]{0pt}{3.5ex} 62 &\hspace{8 mm} 624.8999675(19)& \vline &\hspace{4mm} 100&\hspace{8mm} 625.4561909(15)\\
                   \rule[-1ex]{0pt}{3.5ex}  63 &\hspace{8 mm} 624.9287382(18)& \vline &\hspace{4mm} 101&\hspace{8mm} 625.4629419(17)\\
                    \rule[-1ex]{0pt}{3.5ex} 64 &\hspace{8 mm} 624.9561467(15) & \vline &\hspace{4mm}102&\hspace{8mm}625.4694909(16)\\
                     \rule[-1ex]{0pt}{3.5ex}65 &\hspace{8 mm} 624.9822714(12) & \vline &\hspace{4mm} 103&\hspace{8mm}625.4758492(19)\\
                    \rule[-1ex]{0pt}{3.5ex} 66 &\hspace{8 mm} 625.0071923(19) & \vline &\hspace{4mm} 104&\hspace{8mm}625.4820219(19)\\
                    \rule[-1ex]{0pt}{3.5ex} 67 &\hspace{8 mm} 625.0309862(14) & \vline &\hspace{4mm} 105&\hspace{8mm}625.4880161(18)\\
                    \rule[-1ex]{0pt}{3.5ex} 68 &\hspace{8 mm} 625.0537143(16)& \vline  &\hspace{4mm} 106&\hspace{8mm}625.4938389(19)\\
                     \rule[-1ex]{0pt}{3.5ex}69 &\hspace{8 mm} 625.0754446(13) & \vline &\hspace{4mm}107&\hspace{8mm} 625.4994998(19)\\
                   \rule[-1ex]{0pt}{3.5ex}  70 &\hspace{8 mm} 625.0962310(15) & \vline &\hspace{4mm}108&\hspace{8mm} 625.5050004(17)\\
                   \rule[-1ex]{0pt}{3.5ex}  71 &\hspace{8 mm} 625.1161302(18) & \vline &\hspace{4mm}109&\hspace{8mm} 625.5103468(18) \\
                    \rule[-1ex]{0pt}{3.5ex} 72 &\hspace{8 mm} 625.1351869(16) & \vline &\hspace{4mm}110&\hspace{8mm}625.5155479(20)\\
                   \rule[-1ex]{0pt}{3.5ex}  73 &\hspace{8 mm} 625.1534520(19) & \vline &\hspace{4mm}111&\hspace{8mm}625.5206091(19)\\
                   \rule[-1ex]{0pt}{3.5ex}  74 &\hspace{8 mm} 625.1709705(16) &  \vline &\hspace{4mm}112&\hspace{8mm} 625.5255309(17)\\
                   \rule[-1ex]{0pt}{3.5ex}  75&\hspace{8 mm}  625.1877812(18) & \vline  &\hspace{4mm}113&\hspace{8mm}625.5303229(18)\\
                   \rule[-1ex]{0pt}{3.5ex} 76 &\hspace{8 mm} 625.2039169(17) & \vline &\hspace{4mm}114&\hspace{8mm} 625.5349875(20) \\
                   \rule[-1ex]{0pt}{3.5ex} 77&\hspace{8 mm}  625.2194203(15)& \vline &\hspace{4mm}115&\hspace{8mm}625.5395286(16)\\
                   \rule[-1ex]{0pt}{3.5ex} 78 &\hspace{8 mm} 625.2343196(13)& \vline &\hspace{4mm} 116&\hspace{8mm}625.5439510(18)\\
                   \rule[-1ex]{0pt}{3.5ex} 79 &\hspace{8 mm} 625.2486464(16) & \vline &\hspace{4mm}117&\hspace{8mm}625.5482607(19)\\
                   \rule[-1ex]{0pt}{3.5ex} 80 &\hspace{8 mm} 625.2624314(18) &  \vline&\hspace{4mm}118&\hspace{8mm} 625.5524599(17) \\
                   \rule[-1ex]{0pt}{3.5ex} 81 &\hspace{8 mm} 625.2757008(19) & \vline &\hspace{4mm} 119&\hspace{8mm} 625.5565519(20) \\
                   \rule[-1ex]{0pt}{3.5ex} 82&\hspace{8 mm} 625.2884784(16) & \vline &\hspace{4mm} 120 &\hspace{8mm}625.5605411(19)\\

\hline 
\end{tabular*} 
}
\label{tab:nD5/2}
\end{table}


\section*{Funding}
Ministry of Electronics and Information Technology, India (4(7)/2020-ITEA); Department of Science and Technology, Ministry of Science and Technology, India (SR/WOS-A/PM-59/2019)

\section*{Acknowledgements}
This work was partially supported by the Ministry of Electronics and Information Technology (MeitY), Government of India, through the Center for Excellence in Quantum Technology, under Grant4(7)/2020-ITEA. S. R acknowledges funding from the Department of Science and Technology, India, via the WOS-A project grant no. SR/WOS-A/PM-59/2019. We thank Chris Greene for helpful discussions. We thank Sukanya Mahapatra for her contributions at the early stage of this work; We acknowledge the contribution of Hema Ramachandran, Meena M. S. and the RRI mechanical workshop for the instruments and assistance with the experiments.

\bibliography{ref}

\end{document}